\documentclass[aps,superscriptaddress,amsmath,amssymb,twocolumn,showpacs,floatfix,reprint]{revtex4-1}

\usepackage{graphicx}
\usepackage{amssymb}
\usepackage{bm}
\usepackage{braket}
\usepackage{amsmath}
\usepackage{tabularx}
\usepackage{ltablex}
\usepackage{multirow}
\usepackage[dvipsnames]{xcolor}
\newcommand{\trieste}{Dipartimento di Fisica, Universit\`a di Trieste, Strada Costiera 11, I-34151 Trieste, Italy}
\usepackage{hyperref}
\hypersetup{colorlinks=true, urlcolor=blue, linkcolor=blue, citecolor=blue, pdftex}
\usepackage{verbatim}
\usepackage{float}

\AtBeginDocument{\usepackage{booktabs}}

\begin{document}
\title{Single-point spin Chern number in a supercell framework}

\date{\today}

\author{Roberta Favata}
\affiliation{\trieste}
\author{Antimo Marrazzo}
\email{antimo.marrazzo@units.it}
\affiliation{\trieste}
\begin{abstract}
    We present an approach for the calculation of the $\mathbb{Z}_2$ topological invariant in non-crystalline two-dimensional quantum spin Hall insulators. While topological invariants were originally mathematically introduced for crystalline periodic systems, and crucially hinge on tracking the evolution of occupied states through the Brillouin zone, the introduction of disorder or dynamical effects can break the translational symmetry and imply the use of larger simulation cells, where the $\bf{k}-$point sampling is typically reduced to the single $\Gamma$-point. Here, we introduce a single-point formula for the spin Chern number that enables to adopt the supercell framework, where a single Hamiltonian diagonalisation is performed. Inspired by the work of E. Prodan [Phys. Rev. B, \textbf{80}, 12 (2009)], our single-point approach allows to calculate the spin Chern number even when the spin operator $\hat{s}_z$ does not commute with the Hamiltonian, as in the presence of Rashba spin-orbit coupling. We validate our method on the Kane-Mele model, both pristine and in the presence of Anderson disorder. Finally, we investigate the disorder-driven transition from the trivial phase to the topological state known as topological Anderson insulator. Beyond disordered systems, our approach is particularly useful to investigate the role of defects, to study topological alloys and in the context of ab-initio molecular dynamics simulations at finite temperature. 
\end{abstract}

\maketitle
\section{Introduction}
Two-dimensional (2D) topological insulators (TI) are materials with an insulating bulk and robust edge states protected by the non-trivial topology of the bulk electronic structure~\cite{bernevig_book_2013,vanderbilt_book_2018}. These systems are discussed through topological invariants, integer quantities which characterise the ground-state wavefunction in the bulk. As long as the topological invariant is non-trivial and, possibly, the symmetries needed to define that topology are preserved, the material is said to be in a topological phase. 
These invariants are geometrical properties of the electronic structure, as they are defined in terms of quantities such as the Berry phase or the Berry curvature, which involve derivatives of the occupied states in reciprocal space with respect to the quasi-momentum $\bf{k}$~\cite{vanderbilt_book_2018}. 
Standard geometrical formulas are usually discretised on a regular mesh of $\bf{k}$-points for numerical implementation. However, most electronic structure calculations for non-crystalline systems are normally performed by diagonalising the Hamiltonian at a single $\bf{k}$-point in a large supercell. Usually the $\Gamma$ point at the center of the Brillouin zone (BZ) is considered, although potentially more efficient choices based on the Baldereschi point~\cite{baldereschi_prb_1973} can be employed. The derivation of single-point formulas for geometrical and topological properties is not at all a trivial task, although successful single-point formalism have been developed for the Berry phase~\cite{resta_review_1994}, the orbital magnetisation and the Chern number~\cite{ceresoli_sp_prb_2007}. 

In this work, we target the calculation of the topological invariant for non-crystalline 2D insulators with time-reversal (TR) symmetry. For these systems, encompassing all non-magnetic 2D materials~\cite{mounet_two-dimensional_2018}, the invariant $\nu$ is a $\mathbb{Z}_2$ number: if $\nu=0$ the topology is trivial, otherwise if $\nu=1$ we have a quantum spin Hall insulator (QSHI), where topologically-protected gapless helical edge states cross the bulk gap~\cite{vanderbilt_book_2018}. Over the years, several methods have been developed to calculate the $\mathbb{Z}_2$ invariant in crystalline systems with periodic boundary conditions (PBCs). 

In the following, we briefly outline some of the most popular and practical methods in the context of electronic structure simulations. If inversion symmetry is present, there is a particularly simple method introduced by Fu and Kane~\cite{fukane_prb_2007}, which requires the knowledge of the parity of the occupied states at the four TR-invariant points in the BZ. In the more general case, the $\mathbb{Z}_2$ invariant can be obtained by tracking the evolution of hermaphrodite~\cite{sgiaro_prb_2001} (a.k.a. hybrid) Wannier charge centres~\cite{soluy_prb_2011,soluy_prb_2012,gresch_prb_2017}, or equivalently the eigenvalues of the Wilson loop~\cite{yu_prb_2011,alex_prb_2014,alex_prb_2016}, over half BZ. More recently, a generalisation of the Fu-Kane approach based on elementary band representations~\cite{zak_prb_1982,bradlyn_nature_2017} has allowed to calculate the invariant by using only the knowledge of the irreducible representations of the occupied states at selected high-symmetry points in the BZ~\cite{bradlyn_nature_2017}. 
The $\mathbb{Z}_2$ invariant can be also computed as an individual Chern number~\cite{vanderbilt_book_2018} on half of the Hilbert space~\cite{soluy_prb_2012,gresch_prb_2017}, where the split is performed by two projectors which are smooth and related by TR symmetry.
Although several formulas to compute the $\mathbb{Z}_2$ invariant have been introduced, all the ones we mentioned, and most other existing approaches, require the knowledge of the occupied states at multiple $\bf{k}$-points and become ill-defined for non-crystalline systems; hence in the supercell framework they are of no avail.

Nonetheless, a number of methods have been proposed to deal with non-periodic systems.
Some of these~\cite{essin_moore,leung_prodan,guo} calculate the $\mathbb{Z}_2$ invariant by means of a Pfaffian with twisted boundary conditions, as firstly advocated by Kane and Mele in their original discussion of the $\mathbb{Z}_2$ invariant in presence of disorder and electron-electron interactions~\cite{kane_$z_2$_2005}. A different method is based on constructing the $\mathbb{Z}_2$ invariant from the scattering matrix of the system at the Fermi level~\cite{scattering_insulators,scattering_wire}.
Further, there exists a formulation based on the non-commutative index theorem~\cite{Avron_charge,Avron_index}, where the $\mathbb{Z}_2$ index for disordered topological insulators is computed from the discrete spectrum of a certain compact operator, which is defined as the difference of a proper pair of projection operators~\cite{Katsura_2016, Katsura_2018, Akagi}. 
An alternative non-commmutative approach was proposed by Loring and Hastings~\cite{Hastings_Loring_maths, Hastings_Loring_annals} and relates the $\mathbb{Z}_2$ index to the topological obstruction to approximating almost commuting matrices by exactly commuting matrices; its robustness with respect to the introduction of disorder has been investigated in Ref.~\cite{Loring_disorder}.
The most practical approach from the point of electronic structure simulations has been arguably put forward by Huang and Liu~\cite{huang_prl_2018,huang_prb_2018}, who addressed the problem of calculating the $\mathbb{Z}_2$ invariant for non-periodic system in the context of quantum spin Hall quasicrystals, and introduced the spin Bott index, which measures the commutativity of the projected position operators.
The connection between the Bott indices and Chern or $\mathbb{Z}_2$ invariants has been investigated theoretically~\cite{Toniolo_2022,Loring_disorder,Hastings_Loring_annals,Hastings_Loring_maths}, while numerical simulations~\cite{huang_prb_2018,bott_amorphous,bott_fractal} provided evidence that Bott indices can be used to study non-periodic topological systems. Still, it is conceptually rather unsatisfactory that the calculation of topological invariants in a supercell framework requires introducing radically different formalisms, which call for rather non-trivial equivalence proofs and extensive testing. As a matter of fact, the use of the primitive cell and $\bf{k}$-points is an arbitrary---although indeed very convenient---choice; there is no conceptual reason preventing bona fide $\mathbb{Z}_2$ invariants to be calculated directly in the supercell by deriving a suitable single-point limit. In addition, it is important to assess the convergence with respect to the system size, as different approaches might deliver the same correct answer at very different computational costs.
For instance, recent works ~\cite{huang_prb_2018,Toniolo_2022} claimed that the difference between the Chern number and Bott index is within a correction of the order $O(1/L$), where $L$ is the linear size of the system. Such slow convergence can hinder the study of the system close to a topological phase transition; in fact Huang and Liu empirically added a singular value decomposition (SVD) to their algorithm to improve an otherwise slow convergence~\cite{huang_prb_2018}.

Here, we take a different approach, that essentially combines the work of Ceresoli and Resta on the single-point Chern number~\cite{ceresoli_sp_prb_2007} and the insights from Prodan on a generalised spin Chern number~\cite{prodan_prb_2009}. Notably, our single-point invariant is directly derived by its parent formula for crystalline systems, it shows exponential convergence with the supercell size, both in the pristine and disordered case, it is easy to implement in electronic structure codes, and it works well also in presence of strong Rashba spin-orbit coupling (SOC).
\section{Methods}
In absence of spin-mixing spin-orbit interactions, the spin operator $\hat{s}_z$ commutes with the Hamiltonian and it is possible to discuss the $\mathbb{Z}_2$ invariant in terms of the spin Chern number. In this case, the occupied states diagonalise $\hat{s}_z$ and can be divided in two subsets, either purely spin-up or spin-down, and the regular Chern number can be calculated for each spin. As soon as the Hamiltonian does not commute any more with $\hat{s}_z$, for instance because Rashba SOC is present, such simple-minded spin Chern number cannot be defined any more. Notably, Prodan has shown~\cite{prodan_prb_2009} that it is possible to generalise this definition by projecting the spin operator on the occupied states:
\begin{equation}
	P_z = P(\mathbf{k}) \hat{s}_z P(\mathbf{k}),
\end{equation}
where $P$ is the ground-state projector
\begin{equation}
	P(\mathbf{k}) = \sum_{n} \ket{u_{n\mathbf{k}}}\bra{u_{n\mathbf{k}}},
\end{equation}
$u_{n\mathbf{k}}$ are the periodic part of the Bloch eigenstates and $n$ labels the occupied state at each $\bf{k}$-point in the BZ. Then, we diagonalise $P_z$:
\begin{equation}
	P_z \ket{u_{\lambda}} = s_{\lambda}\ket{u_{\lambda}}.
\label{eq:pz_eigen}
\end{equation}
If only diagonal SOC terms are present, the eigenvalue spectrum of $P_z$ consists of two values only $s_{\lambda}=\pm\frac{1}{2}$ and one can select a single spin component by choosing the eigenstates which correspond to one of the two eigenvalues $s_{\lambda}$.
The crucial observation made by Prodan~\cite{prodan_prb_2009} is that, even if Rashba SOC is present, the spectrum of $P_z$ displays two separate bands  of eigenvalues symmetric around the origin and one can still introduce a well-defined spin Chern number by selecting the eigenvectors with positive (or negative) eigenvalues. Finally, the spin Chern number can be computed as:
\begin{equation}
	C_s = \frac{C_{+}-C_{-}}{2} \textrm{ mod } 2
	\label{eq:spcn}
\end{equation}
where $C_{\pm}$ are calculated on the $u_{\lambda}$ eigenstates with positive and negative eigenvalues respectively; in general it is sufficient to compute either $C_{+}$ or $C_{-}$ only and consider its parity.
The results are of paramount practical relevance, as it is typically much simpler to deal with a formulation based on generalised Chern numbers, which can be written as full BZ integrals and do not require taking into account TR symmetry or complex gauge fixing, as required instead by more general $\mathbb{Z}_2$ formulations~\cite{kane_$z_2$_2005,fukane_prb_2006}.

In principle, if the Rashba interaction is strong enough then the gap of the $P_z$ spectrum might close, preventing the spin Chern number to be defined.
Remarkably, as we will discuss in full detail in the Sec.~\ref{sec:results}, this does not seem to occur in practice. As long as the system is insulating, Eq.~\ref{eq:spcn} is well defined even if the Rashba SOC is several times larger than the diagonal SOC. Hence, we adopt the approach of Prodan~\cite{prodan_prb_2009} and target the derivation of a single-point formula. In order to obtain the correct single-point limit, we follow the approach of Ceresoli and Resta~\cite{ceresoli_sp_prb_2007} for the derivation of the single-point Chern number in TR-broken systems (the latter admit a $\mathbb{Z}$ topological invariant). Let us start with the formula for the generalised spin Chern number in 2D periodic systems:
\begin{eqnarray}
	C_{\sigma} &=& \frac{1}{2\pi} \int_{BZ}Tr_{\sigma}\Omega_{xy}(\mathbf{k}) d\mathbf{k} \nonumber \\
	&=& - \frac{1}{\pi} \sum_{s_{\lambda}=\sigma}\int_{BZ}\mathrm{Im}\braket{\partial_{k_x}u_{\lambda}(\mathbf{k})|\partial_{k_y}u_{\lambda}(\mathbf{k})} dk_x dk_y, \nonumber \\
\end{eqnarray}
where $u_{\lambda}$ are the eigenvectors of $P_z$ (see Eq.~\ref{eq:pz_eigen}) and $\sigma=\pm$ corresponds to one of the sectors of the $P_z$ spectrum. Now we consider the parallelogram Brillouin zone and change coordinate system to have a rectangular integration domain:
\begin{eqnarray}
	C_{\sigma} &=&  -\frac{1}{\pi} \mathrm{Im} \sum_{s_{\lambda}=\sigma} \int_0^{\mathbf{b}_1}d k_1 \int_0^{\mathbf{b}_2} d{k}_2 \braket{\partial_{{k}_1}u_{\lambda}(\mathbf{k})|\partial_{{k}_2}u_\lambda(\mathbf{k})}  \nonumber \\
	&\simeq& -\frac{|\mathbf{b}_1||\mathbf{b}_2|}{\pi}\mathrm{Im} \sum_{s_{\lambda}=\sigma} \braket{\partial_{{k}_1}u_{\lambda}(\mathbf{k})|\partial_{{k}_2}u_\lambda(\mathbf{k})}|_{{\mathbf{k}}=\Gamma},
\end{eqnarray}
where $\mathbf{b}_{1,2}$ are the two reciprocal lattice vectors and the last step is performed in the limit of a very large supercell. In the same limit, we can calculate derivatives through finite differences:
\begin{equation}
	\partial_{{k}_j}\ket{u_\lambda(\mathbf{k})}|_{{\mathbf{k}}=\Gamma} = \lim_{\eta\rightarrow 0 } \frac{\ket{u_{\lambda}(\eta\mathbf{b}_j)}- \ket{u_{\lambda}(\Gamma)}}{\eta|\mathbf{b}_j|},
	\label{eq:fin_diff}
\end{equation}
where we can drop the limit for a large supercell and just consider the difference $\ket{u_{\lambda}(\mathbf{b}_j)}-\ket{u_{\lambda}(\Gamma)}$. 
Eq.~\ref{eq:fin_diff} requires a differentiable function, which is not guaranteed in numerical diagonalisations. Hence, we fix the gauge by adopting a discretised version of the covariant derivative~\cite{sai_prb_2002,ivo_prb_2004} as successfully performed for the Chern number by Ceresoli and Resta~\cite{ceresoli_sp_prb_2007}. One replaces the states with their ``duals'':
\begin{equation}
	\ket{\tilde{u}_{n}(\mathbf{b}_j)} = \sum_{m} S_{mn}^{-1}(\mathbf{b}_j) \ket{{u}_{m}(\mathbf{b}_j)}
\end{equation}
where we define the overlap matrix $S_{nm}(\mathbf{b}_j) = \braket{u_{n}(\Gamma)|u_{m}(\mathbf{b}_j)}$ and the dual states satisfy $\braket{{u}_{n}(\Gamma)|\tilde{u}_{m}(\mathbf{b}_j)}=\delta_{nm}$. Next, we construct the states $u_{n}(\mathbf{b}_j)$ by imposing the periodic gauge, which allows us to perform a single diagonalisation at $\Gamma$:
\begin{equation}
	\ket{u_\lambda(\mathbf{b}_j)} = e^{-i\mathbf{b}_j\cdot\mathbf{r}} \ket{u_\lambda(\Gamma)}.
	\label{eq:periodic_gauge}
\end{equation}
The states in Eq.~\ref{eq:periodic_gauge} are Hamiltonian eigenstates, but they might correspond to a different eigenvalue with respect to the one at $\Gamma$; the ordering is anyway fixed by the covariant derivative.
We note in passing, that while a non-trivial Chern number would prevent the adoption of a periodic gauge for the wavefunction, here the periodic gauge is only temporarily imposed to build each $\ket{{u}_{n}(\mathbf{b}_j) }$ from the knowledge of the $\ket{{u}_{n}(\Gamma) }$, but it is effectively replaced by the parallel transport gauge enforced by the covariant derivative. The final single-point formula for the spin Chern number is
\begin{equation}
	C_{\sigma}^{(asym)} = -\frac{|\mathbf{b}_1||\mathbf{b}_2|}{\pi}\mathrm{Im} \sum_{s_{\lambda}=\sigma} \braket{\tilde{u}_{\lambda}(\mathbf{b}_1)|\tilde{u}_\lambda(\mathbf{b}_2)}.
	\label{eq:spcn_asym}
\end{equation}
In Eq.~\ref{eq:spcn_asym}, we emphasise with the superscript ``asym'' the implicit choice made in Eq.~\ref{eq:fin_diff}, which corresponds to the right-hand derivative. In fact, an alternative choice is the symmetric derivative
\begin{equation}
	\partial_{{k}_j}\ket{u_\lambda(\mathbf{k})}|_{{\mathbf{k}}=\Gamma} \simeq \frac{\ket{u_{\lambda}(\mathbf{b}_j)}- \ket{u_{\lambda}(-\mathbf{b}_j)}}{2|\mathbf{b}_j|},
	\label{eq:sym_dev}
\end{equation}
which can also be computed with a single $\Gamma$-only diagonalisation and leads to the following formula for the spin Chern number:
\begin{widetext}
	\begin{equation}
		C_{\sigma}^{(sym)} = -\frac{|\mathbf{b}_1||\mathbf{b}_2|}{4\pi}\mathrm{Im} \sum_{s_{\lambda}=\sigma} \left(\bra{\tilde{u}_{\lambda}(\mathbf{b}_1)}-\bra{\tilde{u}_{\lambda}(-\mathbf{b}_1)}\right)\left(\ket{\tilde{u}_\lambda(\mathbf{b}_2)}-\ket{\tilde{u}_\lambda(-\mathbf{b}_2)}\right).
		\label{eq:spcn_sym}
	\end{equation}
\end{widetext}
In Sec.~\ref{sec:results}, we will show how the symmetric formula converges much faster than the asymmetric version, at essentially the same computational cost.

We have implemented the single-point formulas in a dedicated Python package, freely available on GitHub~\footnote{https://github.com/roberta-favata/spinv}. The code provides user-friendly interfaces to two popular tight-binding packages such as PythTB~\footnote{https://www.physics.rutgers.edu/pythtb/index.html} and TBmodels~\cite{TB}, and it can be easily interfaced to other codes.

\section{\label{sec:results}Numerical results and discussion}
\begin{figure*}[!ht]
	\centering
	\includegraphics[width=1\linewidth]{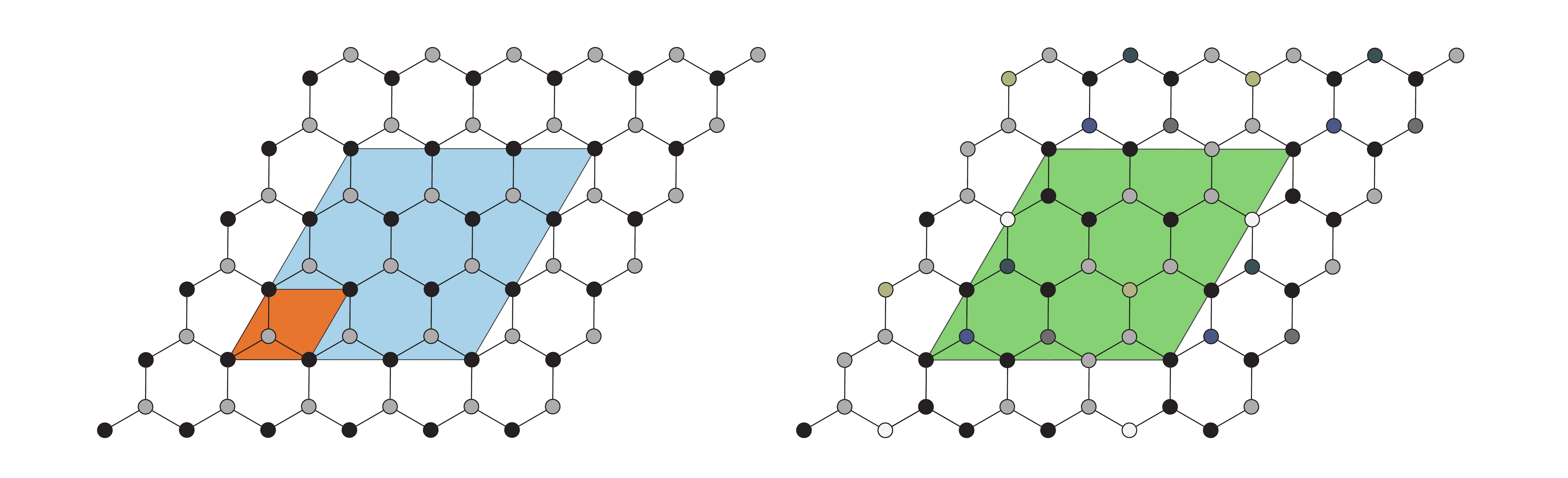}
	\caption{The Kane-Mele model in the supercell approach. Left panel: pristine Kane-Mele model, the primitive cell is shown in orange while a $3\times3$ supercell is marked in blue. Right panel: random realisation of a disordered Kane-Mele model in a $3\times3$ supercell (green) with periodic boundary conditions, where different colours are used to represent the on-site terms. In the following, supercells are labelled by their integer size $L \times L$ (in units of the pristine primitive cell) and the corresponding number of sites $N = 2L^2$.}
	\label{fig:sc}
  \end{figure*}
We validate our approach on the paradigmatic Kane-Mele (KM) model~\cite{kane_quantum_2005,kane_$z_2$_2005} on the honeycomb lattice, both pristine and in presence of Anderson disorder (see Fig.~\ref{fig:sc}). The tight-binding Hamiltonian reads
\begin{eqnarray}
	H_{KM} &=&  t \sum_{ \langle i,j \rangle} c^{\dag}_{i} c^{}_{j} + \Delta \sum_{ i } \xi_i c^{\dag}_{i} c^{}_{i} \nonumber \\
	&+& i \lambda_{SO} \sum_{ \langle \langle i,j \rangle \rangle} \nu_{ij} c^{\dag}_{i }  \sigma^{z}  c^{}_{j } \\ 
	&+& i \lambda_{R} \sum_{ \langle i,j \rangle }  c^{\dag}_{i} ( \boldsymbol{\sigma}\times \hat{\textbf{d}}_{ij} )_{z} c^{}_{j}, \nonumber 
\end{eqnarray}
where $i$ and $j$ run over all sites in the lattice and the creation and annihilation operators are expressed in the contracted form $c^{\dag}_{i} = ( c^{\dag}_{i \uparrow}, c^{\dag}_{i \downarrow} )$.  The first term is a real nearest-neighbour hopping (denoted by $\langle \hspace*{1mm}, \rangle $), if taken alone that would yield four (pair-degenerate) bands with gapless Dirac cones centred on the high-symmetry points $\textbf{K}$ and $\textbf{K}^{'}$ in the Brillouin zone. The second term is a staggered on-site potential ($\xi_i = \pm 1$ is the sublattice index of the $i-$th site) while the third term is the KM SOC~\cite{kane_quantum_2005,kane_$z_2$_2005} which involves a complex next-nearest neighbour hopping (denoted by $\langle \langle \hspace*{1mm}, \rangle \rangle$) with a spin-dependent amplitude proportional to the Pauli matrix $\sigma^{z}$. The factor $\nu_{ij} = \rm{sign} (\textbf{d}_{1} \times \textbf{d}_{2} )_z $ depends on the orientation of the vectors $\textbf{d}_{1}$ and $\textbf{d}_{2}$ along the two bonds connecting $i$ to the next-nearest neighbour site $j$. The fourth term is the Rashba SOC and is a complex nearest-neighbour hopping with off-diagonal spin components, where $\boldsymbol{\sigma} = \left( \sigma^{x},\sigma^{y},\sigma^{z} \right)$ is the vector of Pauli matrices and $\hat{\textbf{d}}_{ij}$ is the unit vector between sites $j$ and $i$.
In the following, we consider a KM Hamiltonian at fixed parameters $t=1$ and $\lambda_{SO}=0.03~t$, which ensure that the energy gap is insulating all over the entire phase diagram~\cite{kane_quantum_2005,kane_$z_2$_2005}.

\subsection{Validation and convergence tests for crystalline systems}

\begin{figure*}[t]
	\centering
	\includegraphics[width=0.4885\linewidth]{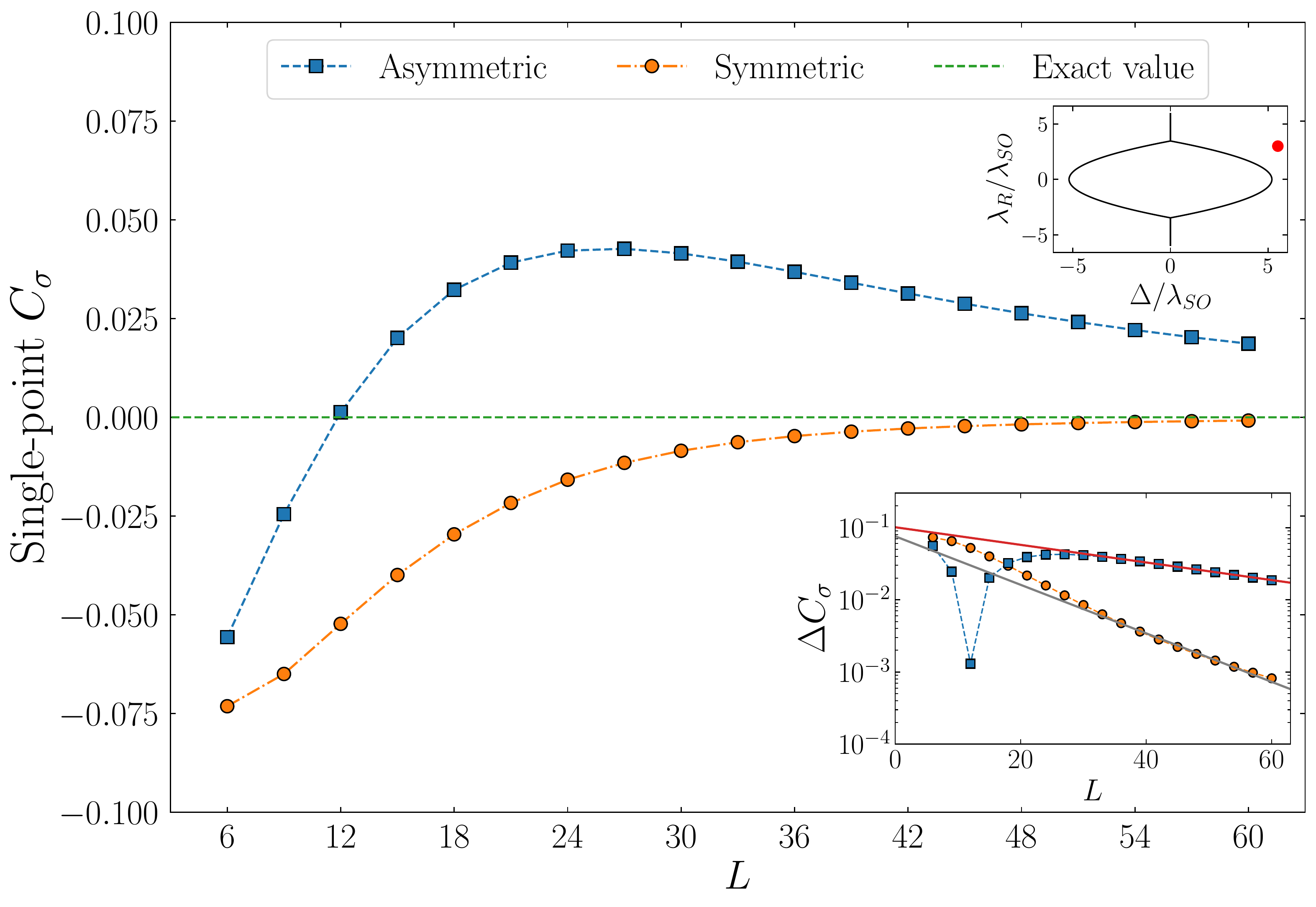}
	\quad \includegraphics[width=0.477\linewidth]{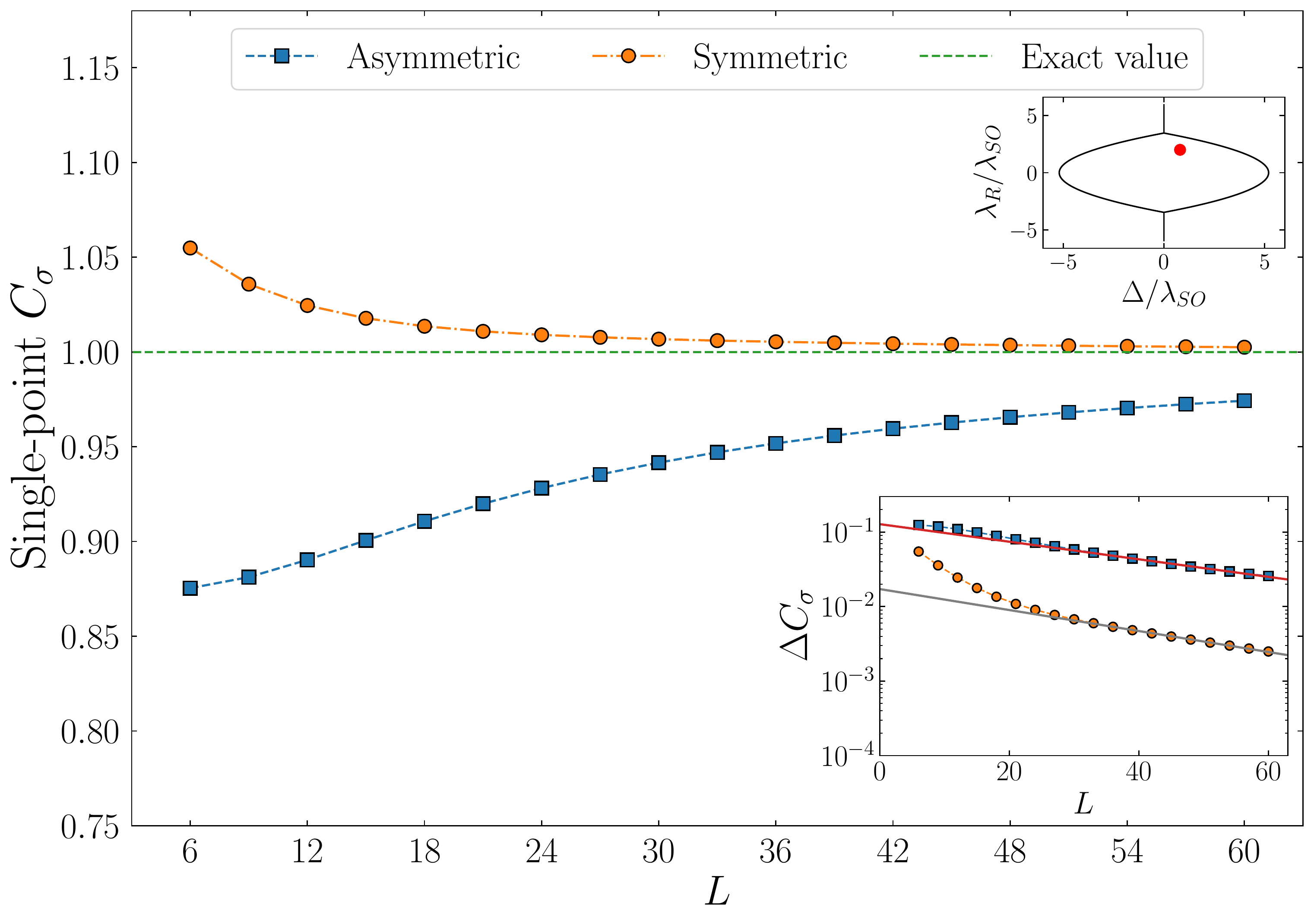}
	\caption{Convergence of the single-point spin Chern number, in its symmetric and asymmetric implementation, with respect to supercell size for the Kane-Mele model, where the Hamiltonian is diagonalised at the $\Gamma$-point only. In the uppest insets, a sketch of the corresponding point in the pristine phase diagram. The lowest insets show the difference between the single-point calculations of the spin Chern number and the thermodynamic limit. Left panel: the spin Chern number converges to zero in the trivial phase ($\Delta/\lambda_{SO} = 5.5$, $\lambda_R/\lambda_{SO}= 3$). Right panel: in the topological phase ($\Delta/\lambda_{SO} = 0.8$ , $\lambda_R/\lambda_{SO} = 2$) the spin Chern number converges to one. In all cases, the asymptotic convergence is exponential, but the symmetric formula converges visibly faster than its asymmetric counterpart.}
	\label{fig:conv_cryst}
\end{figure*}

\begin{figure*}[!ht]
	\centering
	\includegraphics[width=0.63\linewidth]{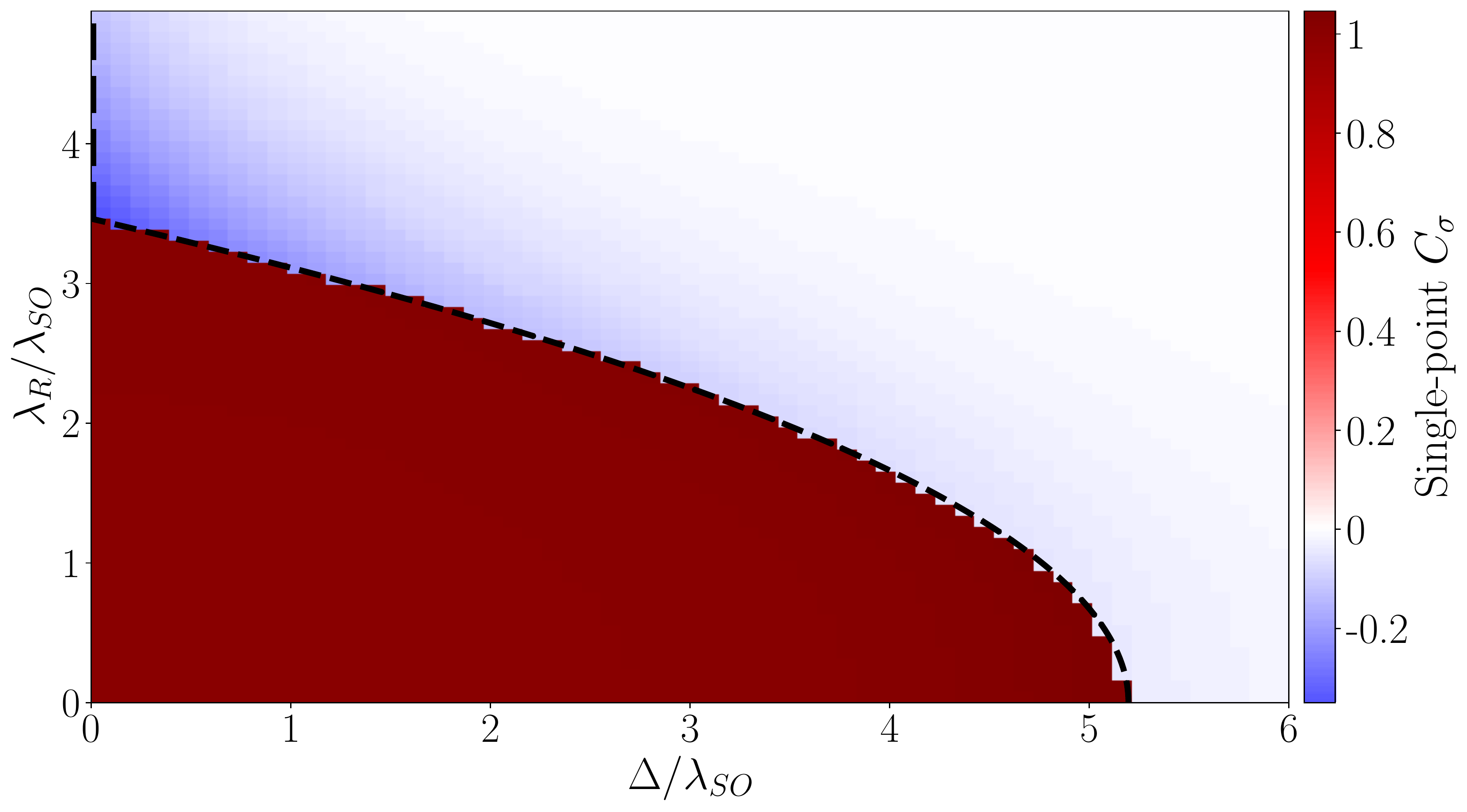} \quad \includegraphics[width=0.532\linewidth]{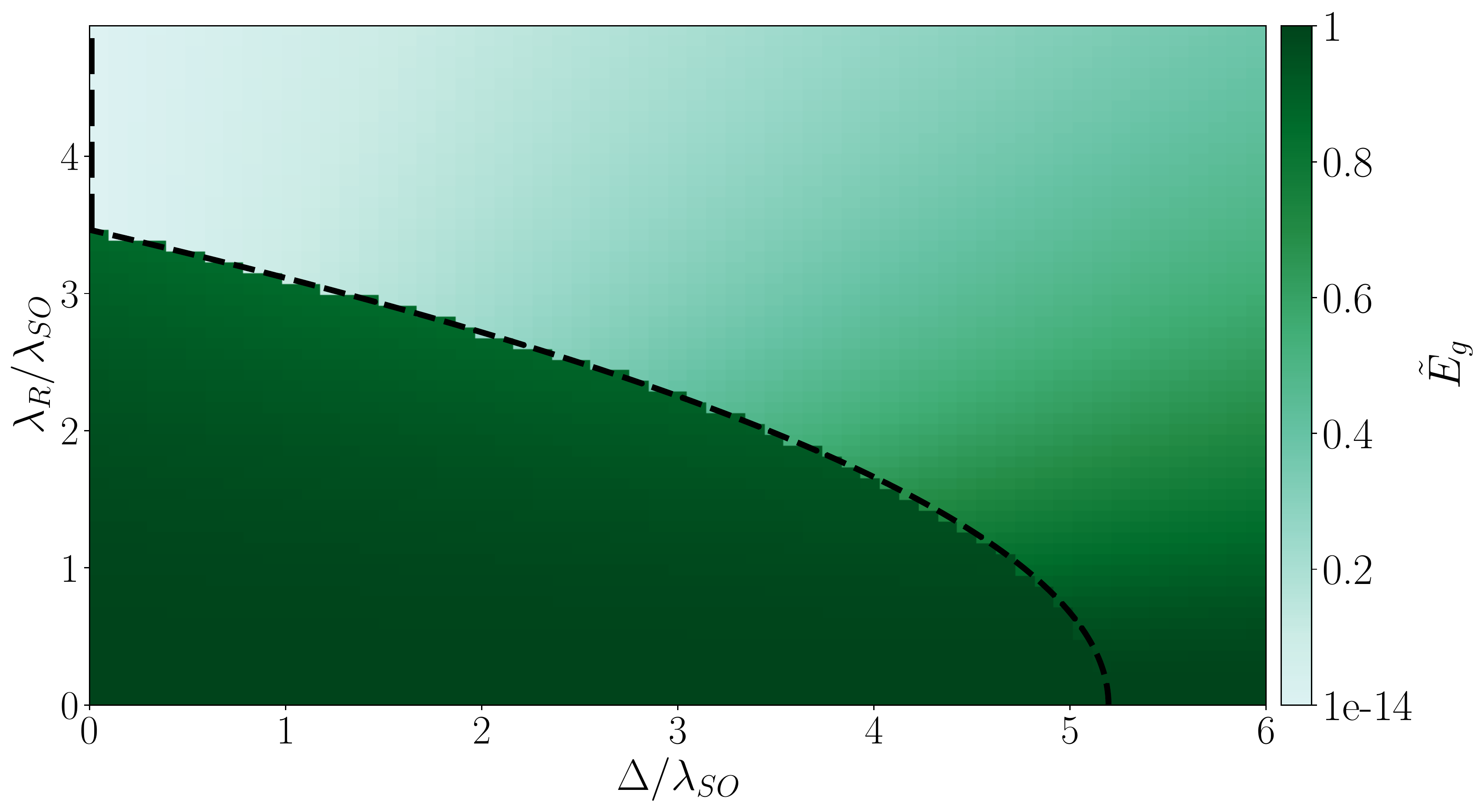} \quad \includegraphics[width=0.433
	\linewidth]{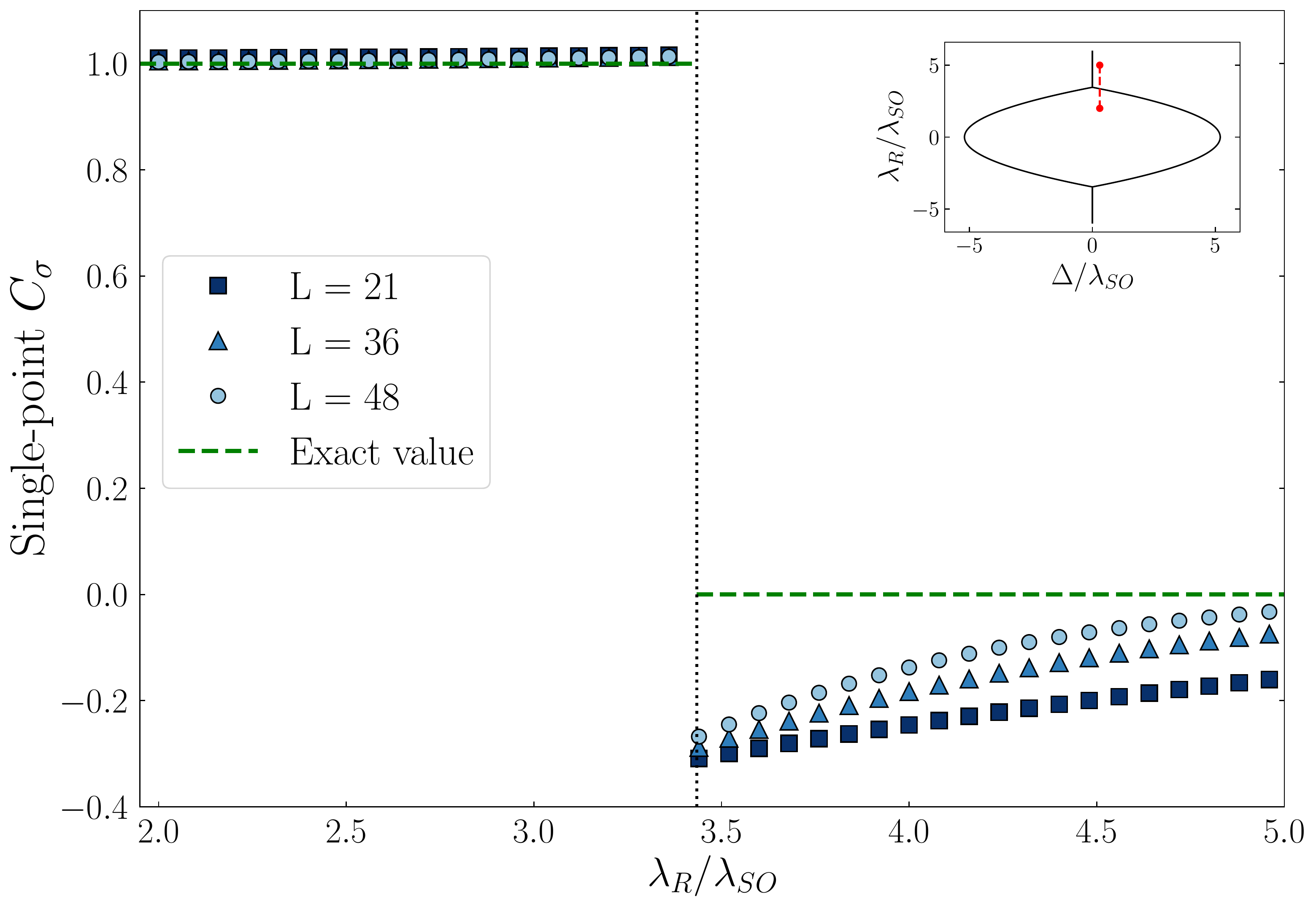}
	\caption{Upper panel: topological phase diagram of the Kane-Mele model calculated with the single-point spin Chern number (symmetric formula), for a supercell size $L=36$ containing $N=2592$ sites. Black dashed line marks the analytical solution for the semi-metallic state separating the topological and trivial phases. Lower-left panel: gap $\tilde{E}_g$ of the $P\hat{s}_zP$ operator for the same calculations performed in the upper panel. Notably, $\tilde{E}_g$ is non-vanishing over all the phase diagram and guarantees that the spin Chern number is well defined everywhere. Lower-right panel: the single-point spin Chern number versus the Rashba coupling $\lambda_R$, at fixed $\Delta/\lambda_{SO}= 0.3$, for different supercell size $L = 21,~36,~48$ and corresponding number of sites $N=882,~2592,~4608$. In that region of the phase diagram, band gaps are very small and finite size effects intensify; still the single-point approach can distinguish the two phases. }
	\label{fig:ph_diag_cryst}
\end{figure*}

In the single-point approach, the topological invariants become exact integer numbers only in the thermodynamic limit of an infinite supercell. First, we test the convergence properties of the single-point spin Chern number (SPSCN) on the pristine KM model, in both asymmetric (Eq.~\ref{eq:spcn_asym}) and symmetric (Eq.~\ref{eq:spcn_sym}) formulation. We inspect the SPSCN as a function of the supercell size $L$, here defined as the number of primitive cells along each lattice vector that makes the supercell ${L}\times {L}$ (see Fig.~\ref{fig:sc}); the number of sites inside the supercell is $N = 2 L^{2}$. A representation of a supercell $3 \times 3$ is given in the left-hand panel of Fig.~\ref{fig:sc}. In our calculations only values of $L$ which are multiple of 3 are considered, to always include the special points $\textbf{K}$ and $\textbf{K}^{'}$ folded at $\Gamma$. We benchmark the accuracy of the formulas inside the $\mathbb{Z}_2$-even and $\mathbb{Z}_2$-odd domains in Fig.~\ref{fig:conv_cryst}. 
\begin{figure*}[t]
	\centering
	\includegraphics[width=0.4875\linewidth]{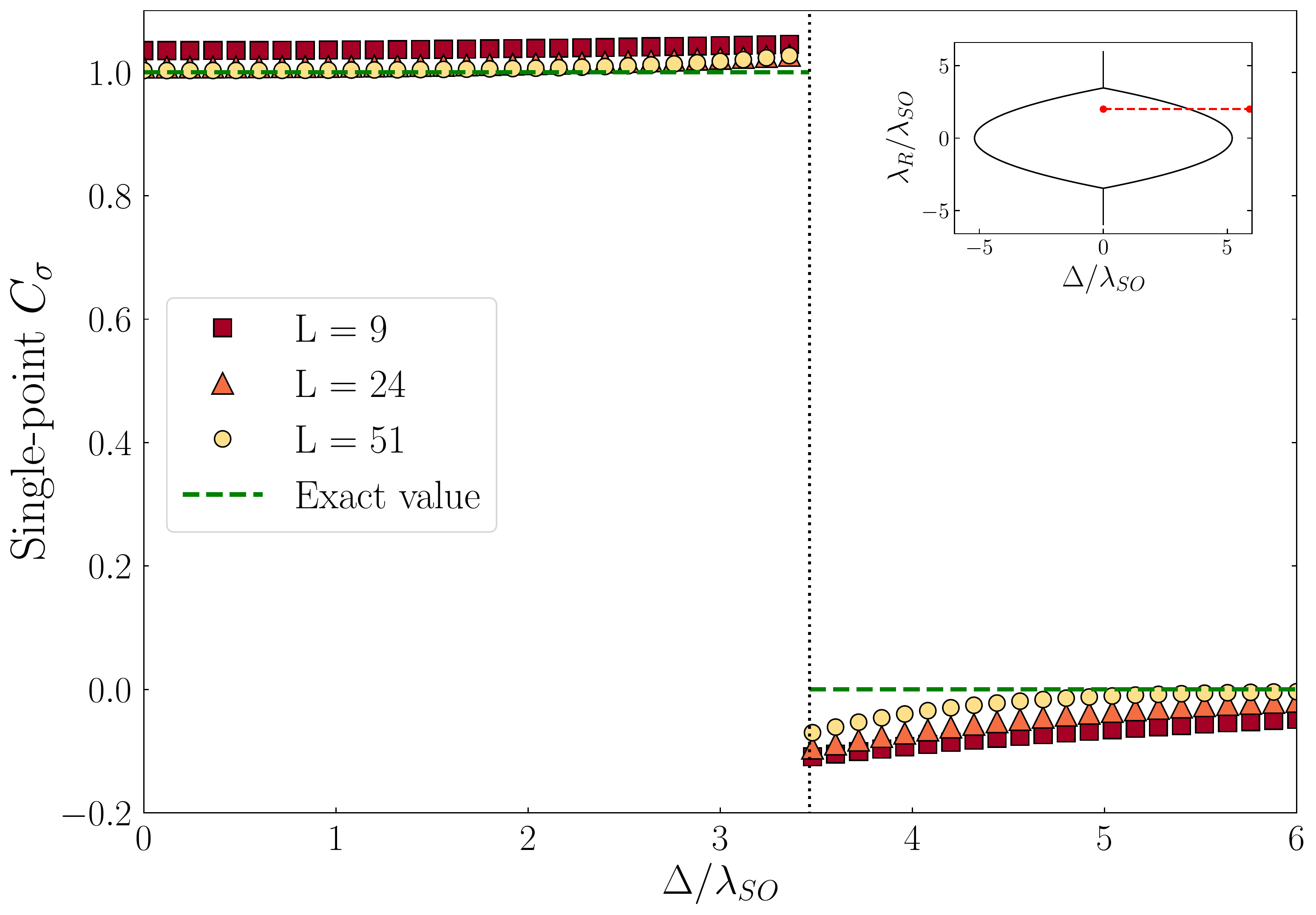} \quad 	\includegraphics[width=0.4783\linewidth]{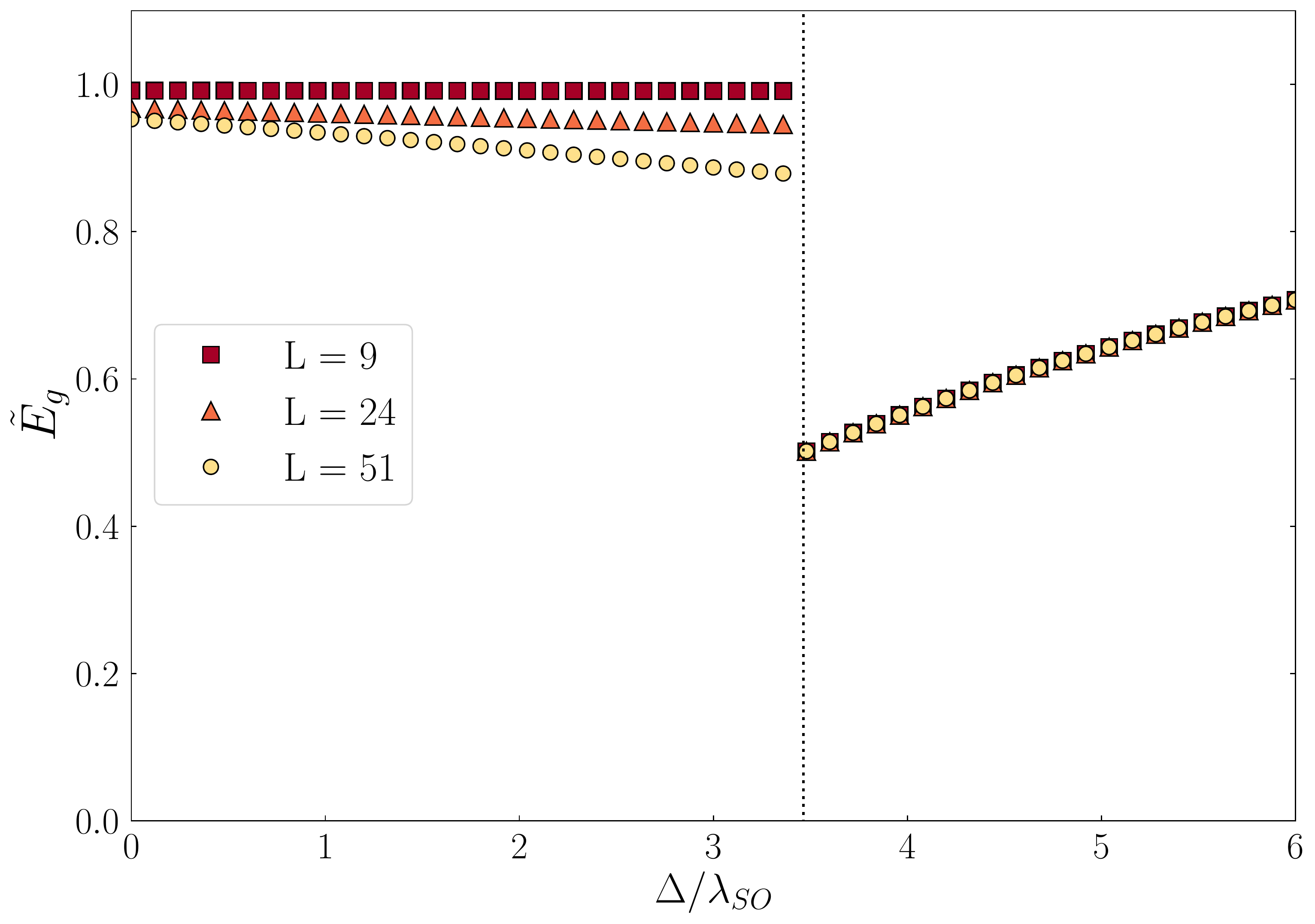}
	\caption{Left panel: the single-point spin Chern number (symmetric formula) versus the on-site term $\Delta$ at fixed $\lambda_{R}/\lambda_{SO} = 2 $ for the Kane-Mele model. Different supercell sizes are considered ($L = 9,~21,~51$, and corresponding number of sites $N = 162,~882,~5202$). As the supercell size increases, the transition becomes sharper and approaches the analytical solution. Right panel: gap $\tilde{E}_g$ of the $P\hat{s}_zP$ operator versus the on-site term $\Delta$ for the same supercell sizes as on the left-hand panel. A non-vanishing $\tilde{E}_g$ guarantees that the spin Chern number is well defined.}
	\label{fig:trans_cryst}
\end{figure*}
The symmetric formula converges faster than the asymmetric one in both trivial and topological phases. Remarkably, the quantity $\Delta C_{\sigma} = | C_{\sigma}(L) - C_{\sigma}(\infty)|$, which is the difference between the spin Chern number given by the single-point formulas at finite sizes and the exact value obtained in the thermodynamic limit, decreases exponentially in both formulations. However, the global prefactor in the symmetric case is an order of magnitude smaller than the one of the asymmetric formula, leading to more accurate results at significantly smaller sizes $L$. Hence, in the following we adopt the symmetric formula only and study the topological phase transition as a function of the on-site $\Delta$, results are reported in Fig.~\ref{fig:trans_cryst}. Our SPSCN is able to reproduce the sharp topological transition already at relatively small supercell sizes, as shown in the left-hand panel of Fig.~\ref{fig:trans_cryst}. The band gap vanishes on the boundary of the phase transition and in the corresponding neighbourhood of parameters convergence is slower and larger supercell sizes must be employed. In the right-hand panel of Fig.~\ref{fig:trans_cryst}, we show how the gap $\tilde{E}_g$ of the $P_{z}$ operator varies across the topological phase transition, but always remains finite, ensuring that our single-point invariant is  everywhere well defined. Then, we validate the SPSCN by calculating the entire topological phase diagram of the KM model, which is reported in the upper panel of Fig.~\ref{fig:ph_diag_cryst}. Notably, the method can distinguish topological and trivial phases even for small, but still finite, values of both the gap of the Hamiltonian and the gap of $P_z$ (lower left-hand panel in Fig.~\ref{fig:ph_diag_cryst}). Larger differences between the SPSCN and the exact value (zero), which are visibile in the upper-left side of the topological phase diagram (marked in blue), are finite size effects and are reduced for large supercells, as highlighted in the lower right-hand panel in Fig.~\ref{fig:ph_diag_cryst}: in that region both to Hamiltonian and $P_z$ operators gap are indeed very small. Therefore, our formulas works well also in presence of very strong Rashba SOC and small band gaps.
\subsection{Disorder-driven topological phase transitions}
The presence of disorder is often modelled by means of an ensemble of large supercells, each representing a specific random realisation as schematically represented in the right-hand panel of Fig.~\ref{fig:sc}. In electronic structure simulations, defect calculations are performed by considering large supercells, to suppress the spurious interactions due to the periodic replicas. Alloys are often simulated through the so-called special quasi-random structures~\cite{zunger_prl_1990}. In addition, a non-perturbative treatment of temperature effects always require working with supercells, being a single structure with special atomic displacements~\cite{zacharias_prr_2020} or a collection of snapshots obtained from ab initio molecular dynamics. 

The SPSCN particularly suits this framework, and we now assess the accuracy and convergence properties of our formula on the KM model supplemented by an Anderson disorder term~\cite{Anderson58}, where we highlight its capability to detect disorder-driven topological transitions. We emphasise that the simple KM model in presence of rather strong Anderson disorder is used as a prototype and a proxy for testing, although our approach targets the more general scenario mentioned above, of supercell calculations, either for model Hamiltonians or first-principles simulations.
\begin{figure*}[ht]
	\centering
	\includegraphics[width=0.4885\linewidth]{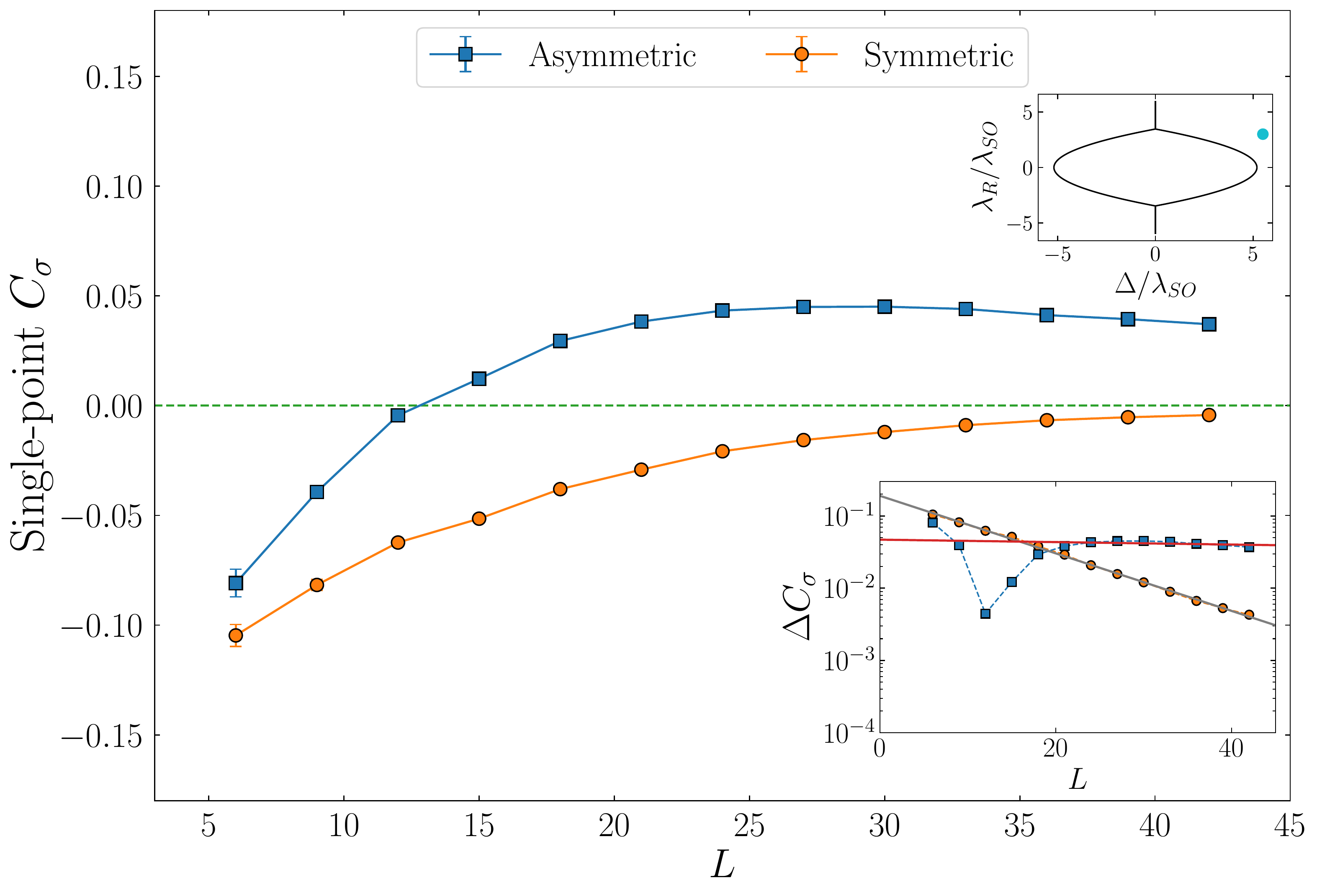}
	\quad \includegraphics[width=0.477\linewidth]{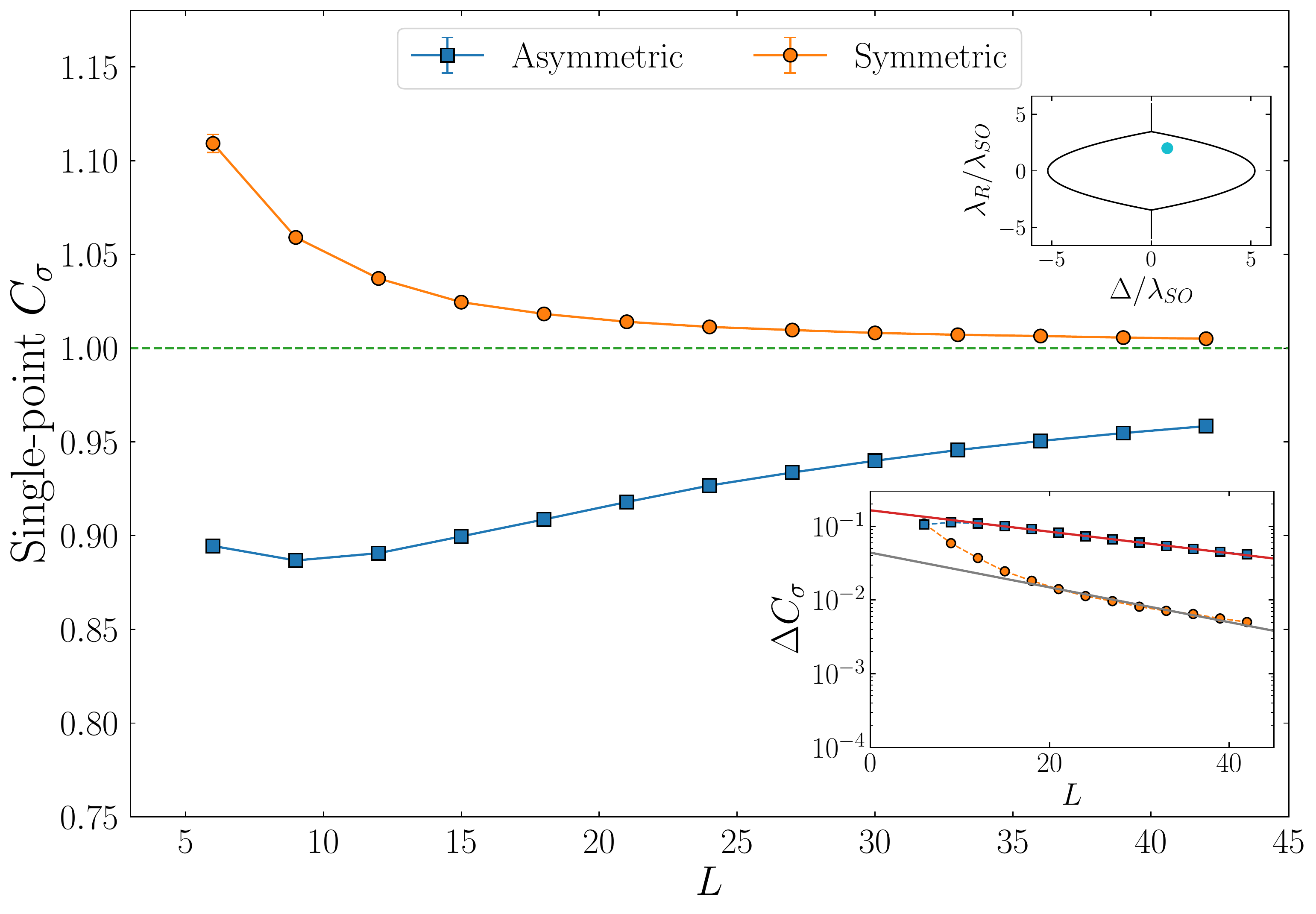}
	\caption{Convergence of the single-point spin Chern number, in its symmetric and asymmetric implementation, with respect to supercell size $L$ for the disordered Kane-Mele model. We report the average and standard deviation of the single-point invariant calculated on $M=100$ realisations with disorder strength $W/t=1$. In the upper insets, the point in the corresponding pristine phase diagram is shown. In the lowest insets, we report the difference between the mean value and the thermodynamic limit as a function of $L$. Left panel: the spin Chern number converges to zero for $\Delta/\lambda_{SO} = 5.5$ and $\lambda_R/\lambda_{SO}= 3$. Right panel: the spin Chern number converges to one for $\Delta/\lambda_{SO} = 0.8$ and $\lambda_R/\lambda_{SO} = 2 $. Also in presence of disorder, the asymptotic convergence is exponential and the symmetric formula converges visibly faster than its asymmetric counterpart. Statistical fluctuations are very small and negligible at almost any supercell size.}
	\label{fig:conv_dis}
\end{figure*}

\begin{figure*}[b]
	\centering
	\includegraphics[width=0.49\linewidth]{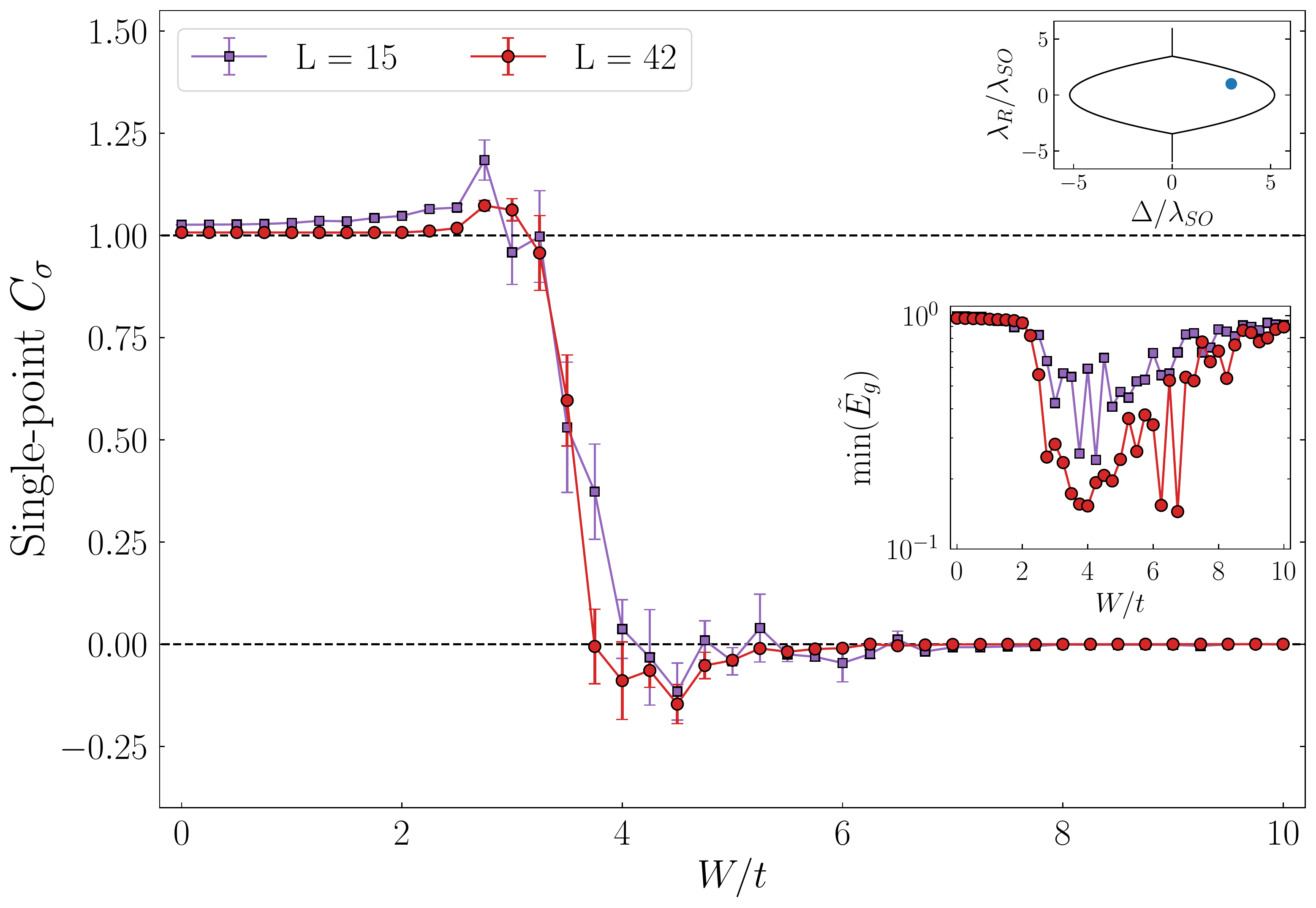}
	\caption{Robustness of the topological phase with respect to disorder. The symmetric single-point spin Chern number is calculated as function of disorder strength $W/t$, starting from the system in the topological phase ($\Delta/\lambda_{SO} = 3$, $\lambda_R/\lambda_{SO} = 1$). For each $W$, we report the mean and standard deviation over $M=50$ realisations of Anderson disorder for supercells of sizes $L=15,~42$ and number of sites $N = 450,~3528$ respectively.  Upper inset: a sketch of the point where the calculations are computed reported on the pristine phase diagram $(W/t=0)$. Lower inset: minimum value, over the disorder realizations, of the gap $\tilde{E}_g$ of the $P\hat{s}_zP$ operator as a function of $W/t$. With increasing supercell size $L$, the transition becomes sharper. $\tilde{E}_g$ does not vanish with Anderson disorder and the approach performs well also in the strong-disorder regime.}
	\label{fig:dis_transition}
\end{figure*}

\begin{figure*}[t]
	\centering
	\includegraphics[width=0.483\linewidth]{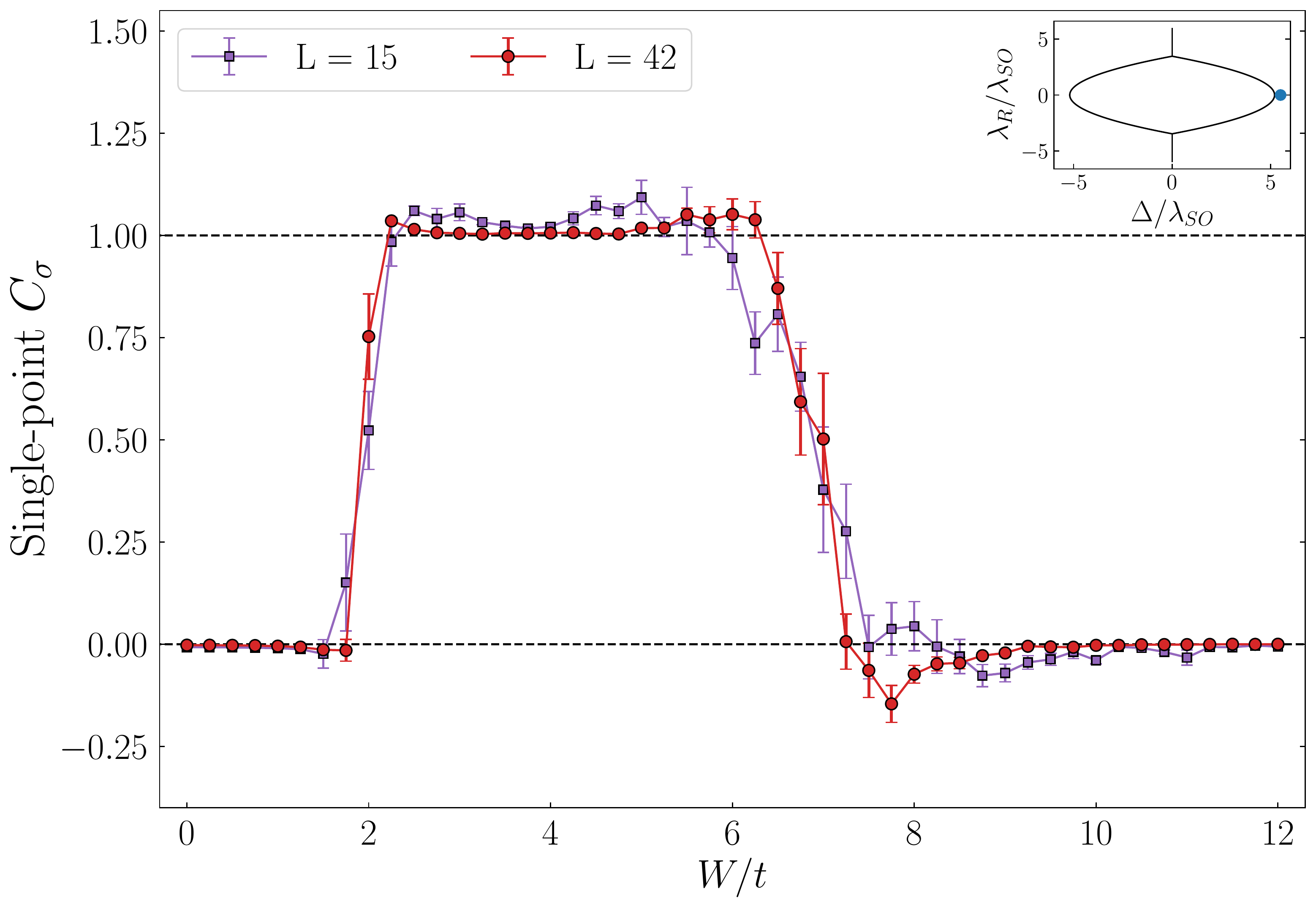} \quad \includegraphics[width=0.482\linewidth]{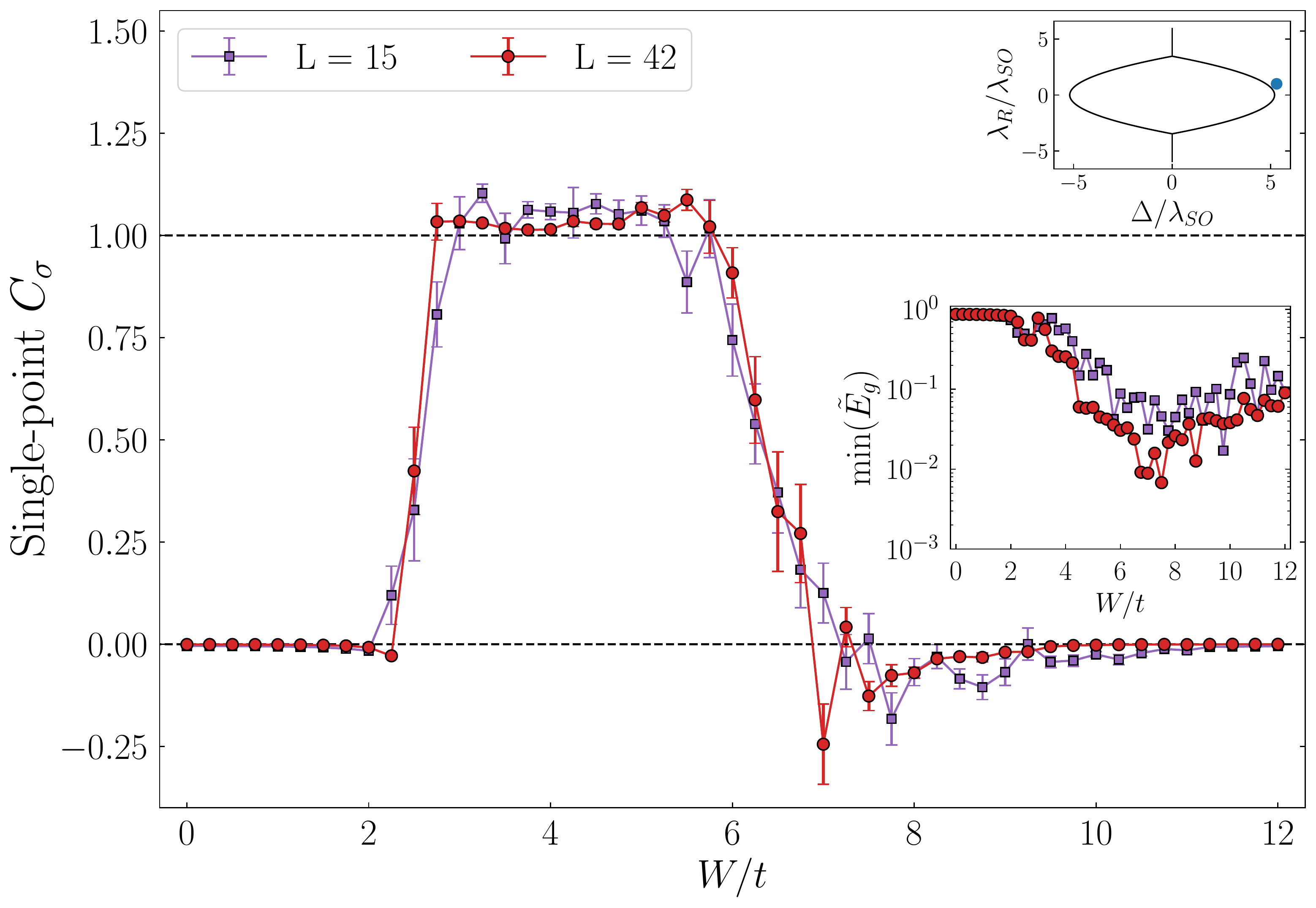}
	\caption{Topological Anderson insulator (TAI). The symmetric single-point spin Chern number is calculated as function of disorder strength $W/t$, starting from the system in a trivial state close to the phase transition. We report the mean value and the standard deviation of single-point invariant over $M=50$ realisations of Anderson disorder for supercells of sizes $L = 15,~42$ and corresponding numbers of sites $N = 450,~3528$ respectively. For $5 \le W/t \le 10$ the number of random realisations is purposely increased to $M=100$ to reduce the standard deviation.  Left panel: TAI state in absence of Rashba coupling ($\Delta/ \lambda_{SO} = 5.5$, $\lambda_R = 0$).
	Right panel: TAI at finite Rashba coupling ($\Delta/ \lambda_{SO} = 5.3$, $\lambda_R/\lambda_{SO}= 1$). Here, the minimum value of the gap $\tilde{E}_g$ (over $M$ disorder realizations) is reported versus $W/t$ in the lower inset (the same plot is not present in the right-hand left panel since $\tilde{E}_g$ is constantly equal to one for $\lambda_R = 0$). }
	 \label{fig:tai}
\end{figure*}

The Hamiltonian of the disordered KM model reads
\begin{equation}
	H_{dis} = H_{KM} + \sum_{i} w_i c^{\dag}_{i} c^{}_{i},
	\label{eq:dis_KM}
\end{equation}
where $w_i \in \left[ -\frac{W}{2},\frac{W}{2} \right] $ is a randomly distributed on-site potential and $W$ is the disorder strength which, in the following, is reported in units of the nearest-neighbour hopping amplitude $t$. In Fig.~\ref{fig:conv_dis} we test the convergence of the single-point formulas (Eqs.~\ref{eq:spcn_asym} and~\ref{eq:spcn_sym}) with increasing supercell size $L$ for the disorder strength $W/t = 1 $, which is weak enough not to destroy the topological phases of the corresponding pristine KM model. The SPSCN is evaluated as the mean value over $M$ realisations of random disorder with supercells of size $L \times L$. Also in presence of disorder, the convergence of the formulas is exponential and the symmetric version converges faster than the asymmetric one. In addition, we consider increasing disorder strengths and study the robustness of the topological phase, results are reported in Fig.~\ref{fig:dis_transition}. For sufficiently strong disorder, the topological phase is destroyed and the SPSCN becomes trivial. As expected, the width of the phase transition becomes smaller with increasing supercell sizes.
As investigated in Ref.~\cite{KM_anderson}, for a certain range of parameters, the disordered KM model given by Eq.~\ref{eq:dis_KM} displays a topological state called topological Anderson insulator (TAI). It is a phase of quantized conductance which is obtained adding Anderson disorder to a trivial insulator or metal which are relatively close to a topological phase transition~\cite{TAI,TAI_Beenakker,TAI_phenomena}. The mechanism for this disorder-induced transition has been discussed in terms of a renormalization of the model parameters such as the on-site term~\cite{TAI_Beenakker}. The weak-disorder boundary of a TAI can be studied within an effective-medium theory and the self-consistent Born approximation~\cite{TAI_Beenakker,KM_anderson}, but these perturbative approach might fail in the strong-disorder regime, where the TAI phase is destroyed in favour of a trivial insulating phase, as we show next. In Fig.~\ref{fig:tai} we use the SPSCN to inspect these topological phase transitions driven by disorder. In order to compare with previous work on the disordered KM model~\cite{KM_anderson} and for the sake of clarity, we consider a value of $\lambda_{SO} = 0.3~t$ which is an order of magnitude greater than the one used for the previous examples. First, we fix $\lambda_R = 0$ (left-hand panel) and observe that the TAI appears at about $W/t = 2$, in agreement with the conductance calculation in~\cite{KM_anderson} and the spin Bott index results in~\cite{huang_prb_2018} (note the factor of two with respect to the $W$ defined therein). Then, we consider finite Rashba SOC and show the results in the right-hand panel of Fig.~\ref{fig:tai}, where we note that the TAI region has become narrower, in agreement with Ref.~\cite{KM_anderson}. A check on the gap $\tilde{E}_{g}$ of operator $P_z$ is performed for every SPSCN calculation in presence of disorder: Anderson disorder never fully closes the gap and the invariant can always be computed.

\section{Conclusions}

In this work, we have introduced a robust and efficient single-point formula to calculate the $\mathbb{Z}_2$ topological invariant in non-crystalline 2D materials. We have validated our method with supercell numerical simulations on the KM model, both pristine and disordered. Our approach can reproduce the entire phase diagram of the KM model, where each calculation requires only a single-point diagonalisation in the supercell framework, even in presence of strong Rashba SOC. In addition, we have extensively tested our method in presence of Anderson disorder, and we have shown how the single-point formula can correctly describe disorder-driven topological phase transitions. In particular, we have discussed both the process where disorder destroys the topological phase and where disorder actually promotes it, as for the TAI phase; that is in agreement with calculations of the conductance~\cite{scattering_wire,KM_anderson} and spin Bott index~\cite{huang_prb_2018} reported in the literature. Our single-point approach converges exponentially with size, so it is typically sufficient to work with relatively small supercells, which is critical for applications in ab initio modelling. 
One of the side benefits of adopting Prodan's approach is that the formula can, at least in principle, be meaningful also in presence of weak TR-breaking perturbations~\cite{prodan_prb_2009}. This feature could be useful to study how the bulk topology is affected by the presence of magnetic impurities, or of a magnetic substrate through the proximity effect; even though the absence of TR symmetry would allow backscattering between the two helical edge states.
To encourage the use of our approach, we release a dedicated Python package that allows to seamlessly calculate the single-point Chern ($\mathbb{Z}$) and spin-Chern ($\mathbb{Z}_2$) invariants of any TB model thanks to dedicated interfaces to PythTB and TBmodels, two very popular TB codes. Notably, these two packages also allow working with Wannier Hamiltonians, which are read in the format produced by Wannier90~\cite{Mostofi2008,Pizzi2020}; that provides a simple way to apply our work in the context of first-principles calculations. Then, it would be interesting to explore the effect of the TB approximation where the real-space position operator is taken to be diagonal, versus considering all off-diagonal elements, essentially taking into account the overlap between Wannier functions. Nonetheless, the formalism is rather simple and it could be implemented with limited effort directly into plane-wave first-principles codes, such as Quantum ESPRESSO~\cite{Giannozzi2009,Giannozzi2017}. In short, our approach allows studying 2D topological insulators in a supercell framework, which is crucial to investigate very relevant phenomena such as disorder, defects, alloying, and to study dynamical and temperature effects through ab initio molecular dynamics simulations.

\bibliography{biblio}
\end{document}